\documentclass[runningheads]{llncs}
\usepackage[T1]{fontenc}
\usepackage{multirow}
\usepackage{graphicx}
\usepackage{hyperref}
\begin{document}

\title{Developing Authentic Simulated Learners for Mathematics Teacher Learning: Insights from Three Approaches with Large Language Models}

%Simulating Realistic Students for Responsive Teaching: A Comparative Study of LLM Enhancement Techniques.
\author{Jie Cao\inst{1}\orcidID{0009-0002-6507-9613}$^*$ 
Ha Nguyen\inst{1}\orcidID{0000-0001-7138-1427}$^*$ \and
Selim Yavuz\inst{2}\orcidID{0009-0005-6816-2422} \and Boran Yu\inst{1}\orcidID{0000-0002-2212-9038}\and Shuguang Wang\inst{1}\orcidID{0009-0008-6781-2448} \and Pavneet Kaur Bharaj\inst{3}\orcidID{0000-0002-6742-8074} \and Dionne Cross Francis\inst{1}\orcidID{0000-0001-6851-8867}}
\authorrunning{Cao et al.}
\titlerunning{Developing Authentic Simulated Learners}


\institute{The University of North Carolina at Chapel Hill, North Carolina, USA \and Indiana University Bloomington, Indiana, USA \and California State University Long Beach, California, USA}

\maketitle              % typeset the header of the contribution

\def\thefootnote{*}\footnotetext{These authors contributed equally to this work.}\def\thefootnote{\arabic{footnote}}

\begin{abstract}
Large Language Model (LLM) simulations, where LLMs act as students with varying approaches to learning tasks, can support teachers' noticing of student thinking. However, simulations using zero- or few-shot prompting often yield inauthentic knowledge and language, directing teachers to unrealistic reasoning. We evaluate three approaches (Fine-tuning, Multi-agent, and Direct Preference Optimization; DPO) to improve the authenticity and pedagogical utility of simulated students. All approaches improve cognitive and linguistic authenticity, compared with few-shot prompts. Interviews with elementary mathematics pre-service teachers and researchers (\textit{n} = 8) reveal distinct pedagogical affordances. The fine-tuned model produces realistic, brief responses but limits opportunities to extend students' thinking. Meanwhile, the multi-agent and DPO approaches generate explicit reasoning behind student strategies. We discuss implications for designing LLM simulations that balance authenticity with instructional utility for teacher learning.

\keywords{simulations \and teacher noticing \and elementary mathematics \and LLMs \and fine-tuning \and multi-agent \and Direct Preference Optimization}
\end{abstract}
\section{Introduction}
Responsiveness in mathematics education refers to \textit{professional noticing of students' mathematical thinking} with three interconnected skills: attending to students' strategies, interpreting their emerging understanding, and deciding how to respond to build on that understanding \cite{jacobs2010professional}. Developing noticing skills is challenging for pre-service teachers (PSTs), who tend to prioritize guiding students to correct solutions rather than enriching their mathematical interpretations \cite{kilic2022preservice}. To support teacher noticing, practice-based teacher education (PBTE) decomposes noticing practices into component skills and allows teachers to rehearse in low-risk environments \cite{grossman2009teaching}. PSTs can practice teaching interactions in simulated classrooms using human role-playing \cite{lampert2013keeping}, mixed-reality \cite{cohen2020teacher}, and more recent Large Language Models (LLMs) student simulations as a scalable solution \cite{pan2025tutorup,zhang2025seeking}. 

%While researchers use specific prompts to emulate student proficiency and motivation \cite{jin2025teachtune}, LLMs often default to acting like AI assistants or experts rather than real learners \cite{macneil2024synthetic}. Furthermore, they can drift across conversational turns, causing inconsistencies in simulated cognition and language \cite{liu2025llms,martynova2025can,zheng2025teaching}. Such inauthentic reasoning compromises teachers' engagement and the simulation's overall effectiveness \cite{howell2021approximations}. 

Researchers have used specific prompts to emulate students' knowledge and motivation in LLM simulations \cite{jin2025teachtune}. However, LLMs might show characteristics of AI assistants or experts rather than learners \cite{macneil2024synthetic}. They might drift across conversational turns and show inconsistencies in simulating cognition and language \cite{liu2025llms,martynova2025can}, negatively impacting teacher engagement and the simulations' effectiveness. To address these challenges, we explore approaches beyond simple prompting (e.g., zero-shot, few-shot), namely Fine-tuning \cite{xu2025classroom,zheng2025cognitive}, Multi-agent systems \cite{cao2025first}, and Direct Preference Optimization (DPO) \cite{scarlatos2024improving}. We ask: \textbf{RQ1:} \textit{To what extent do Fine-tuning, Multi-agent, and DPO approaches reflect authentic student cognition and language, compared to few-shot prompts?} \textbf{RQ2:} \textit{How do teachers perceive the authenticity and pedagogical utility of the simulated students?}

\section{Related work}
\subsection{Teacher Noticing in Mathematics Learning}
Responsive teaching in mathematical education requires a high degree of \textit{professional noticing of students' mathematical thinking}, including attending to students' reasoning and supporting equitable, student-centered classroom discourse \cite{jacobs2010professional,vanes2021expanding}. Although promising, many elementary teachers and PSTs struggle to enact professional noticing. They only pursue ideas that appear accurate and do not always connect students' ideas and mathematical concepts \cite{mikeska2025promoting}. 

PBTE advances professional noticing by allowing PSTs to rehearse authentic teaching \cite{grossman2009teaching}. Traditional PBTE approaches, such as shadowing practicing teachers and implementing instruction in field experiences, are short and can negatively impact student learning. Teacher educators have implemented alternative practices, including video reflections \cite{richter2022video}, peer rehearsals \cite{Lee02012024}, and role-playing \cite{mikeska2025promoting}. However, video reflections lack the urgency of real-time decision-making, and peer or role-play scenarios may not capture authentic student reasoning.

\subsection{LLM-based Student Simulations}
LLMs-based simulations---where LLMs role-play as students---offer a promising PBTE approach that overcomes these limitations~\cite{Barrett2025,jin2025teachtune,pan2025tutorup}. However, researchers have identified an \textit{authenticity gap}. LLM simulations may inconsistently or inaccurately simulate students' understanding and error patterns~\cite{macneil2024synthetic,zheng2025teaching} and rely on overly verbose and complex language that does not reflect student talk  \cite{martynova2025can}. 

To address these issues, we turn to three approaches: Fine-tuning, Multi-agent architectures, and DPO. \textbf{Fine-tuning} involves training LLMs on domain-specific data to align outputs with reference patterns. PSTs who interacted with LLMs fine-tuned with classroom discourse data reported that the interactions felt naturalistic and positively impacted how they would approach responsive teaching \cite{Barrett2025,zheng2025teaching}. 
We also explore \textbf{Multi-agent architectures}, which employ multiple LLMs to collaborate, critique, and self-correct responses, to improve LLMs' reasoning in complex tasks \cite{li2024survey}. Finally, we use \textbf{Direct Preference Optimization (DPO)}, which provides paired preference data and steers the model's output towards preferred behaviors \cite{rafailov2023direct}. DPO has shown promise in increasing the accuracy and pedagogical alignment of LLM-generated feedback in mathematics education \cite{scarlatos2024improving}. Notably, these approaches differ in data requirements and adaptivity to feedback. To our knowledge, no prior work has systematically compared these approaches to examine the authenticity of student dialogues. Our study thus offers design-relevant evidence for scaling LLMs simulations.

\section{Method}
\subsection{The Teaching Task and Simulated Student Profile}
\begin{figure}
    \centering
    \includegraphics[width=.85\linewidth]{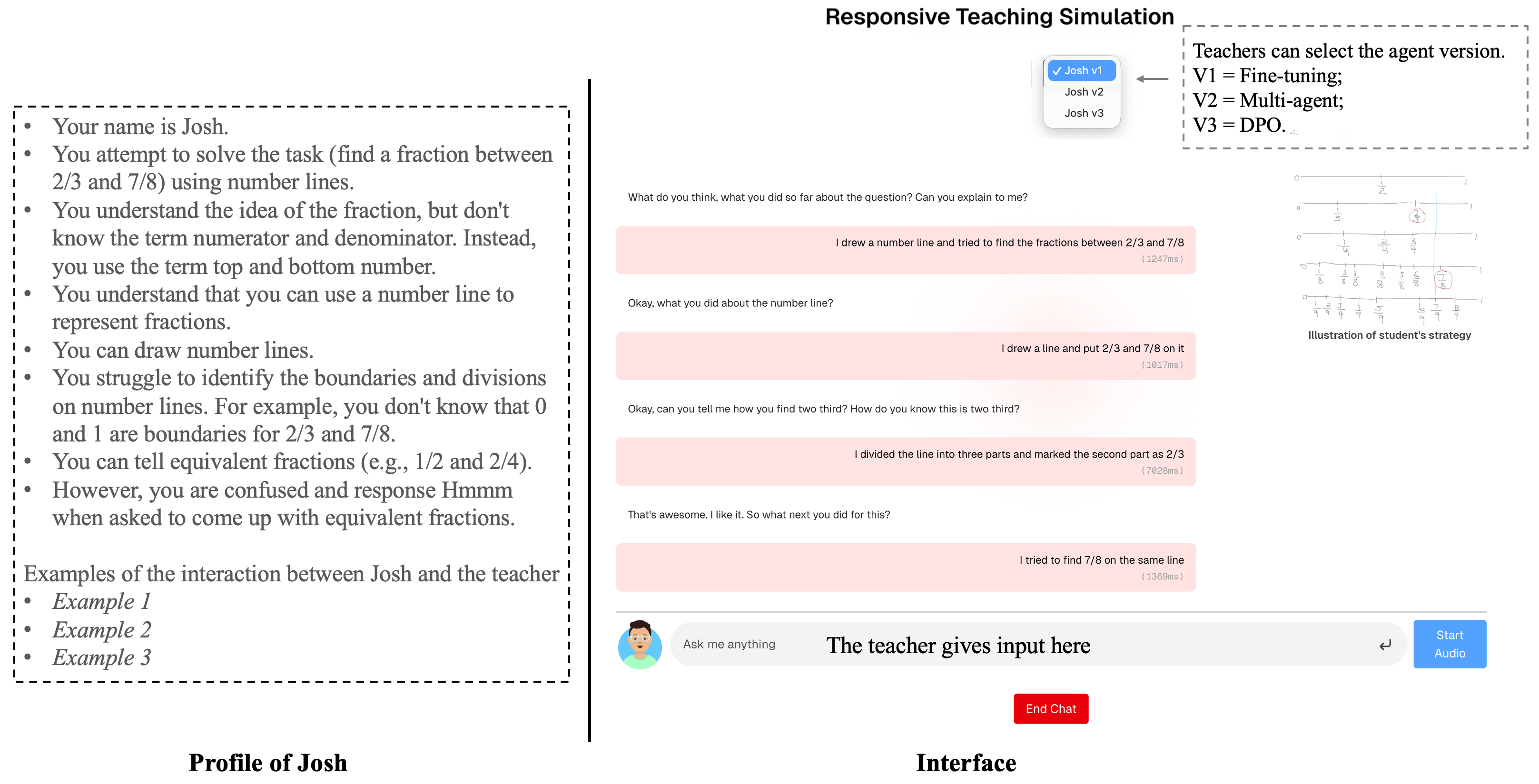}
    \caption{Prompt and interface for the simulated agent (Josh)}
    \label{fig:josh}
\end{figure}

In the simulation (Figure \ref{fig:josh}), PSTs engaged in one-on-one, text or voice interactions with an LLM agent (``Josh''), who role-played as a fifth-grader finding ``a fraction between 2/3 and 7/8.'' This task helped PSTs practice eliciting student thinking in working with fractions, a key topic in elementary classrooms. We implemented the simulation with 15 PSTs in a mathematics teaching methods class at a public university in the United States (Fall 2025). The PSTs interacted with the simulations five times (10-15 minutes/interaction), totaling 1438 talk turns.
We used few-shot prompting in this initial implementation (left; Figure \ref{fig:josh}). PSTs reported that the simulated student sometimes appeared inauthentic (e.g., ``talking too much'', ``being too smart''), which affected their attempts to elicit or extend the agent's thinking. These observations motivated us to identify instances where the agent might appear inauthentic and improve its output. 

%We detailed the task, the student's (Josh) knowledge (e.g., ``you struggle to identify the boundaries on number lines''), and three examples of teacher-student interactions illustrating common student strategies (e.g., drawing several number lines to find the solution). The prompts and examples were drafted by three researchers, who had experience teaching mathematics in K-12, researching mathematics education, and preparing PSTs. To encourage student-like discourse, we added instructions for the agent to keep the responses short, include fillers and uncertainty (e.g., ``eh'', ``hmm'', ``uh''), and reveal its thinking slowly. %when prompted by the teachers.

\subsection{Evaluation Framework of the Simulated Student}
Two researchers with mathematics teaching experience analyzed 20\% of PST-agent interactions from the first implementation. They identified 40 interactions that appeared inauthentic and wrote analytic memos to document their reasoning. We inductively analyzed the memos to develop an evaluation framework, focusing on Cognition and Language (building on \cite{liu2025llms,martynova2025can}). Appendix A provides code descriptions\footnote{\url{https://osf.io/5nv6u/overview?view_only=a0f57da0ddc746c58d1156fccb24d211}}). 
For \textbf{Cognition}, responses should align with the student profile in \textit{knowledge scope} and show \textit{logical consistency} in understanding across turns. It should not express \textit{inconsistent uncertainty}. %That is, if the agent expresses confusion about a concept, it should not fluently explain the same concept in the subsequent turn. 
The \textit{explanation level} should not be too complete to reflect fragmented or emerging understanding of fractions. Regarding \textbf{Language}, responses should reflect a \textit{natural student-like tone}. They should avoid \textit{formal} language (using disciplinary terminologies) and \textit{formulaic} structure (repeating scripted phrases). %Together, these dimensions align with prior research that assessed the cognitive and language complexity of LLM-simulated students \cite{liu2025llms,martynova2025can}. 

\subsection{Three Approaches for Developing the Simulated Student}
\begin{figure}
    \centering
    \includegraphics[width=.7\linewidth]{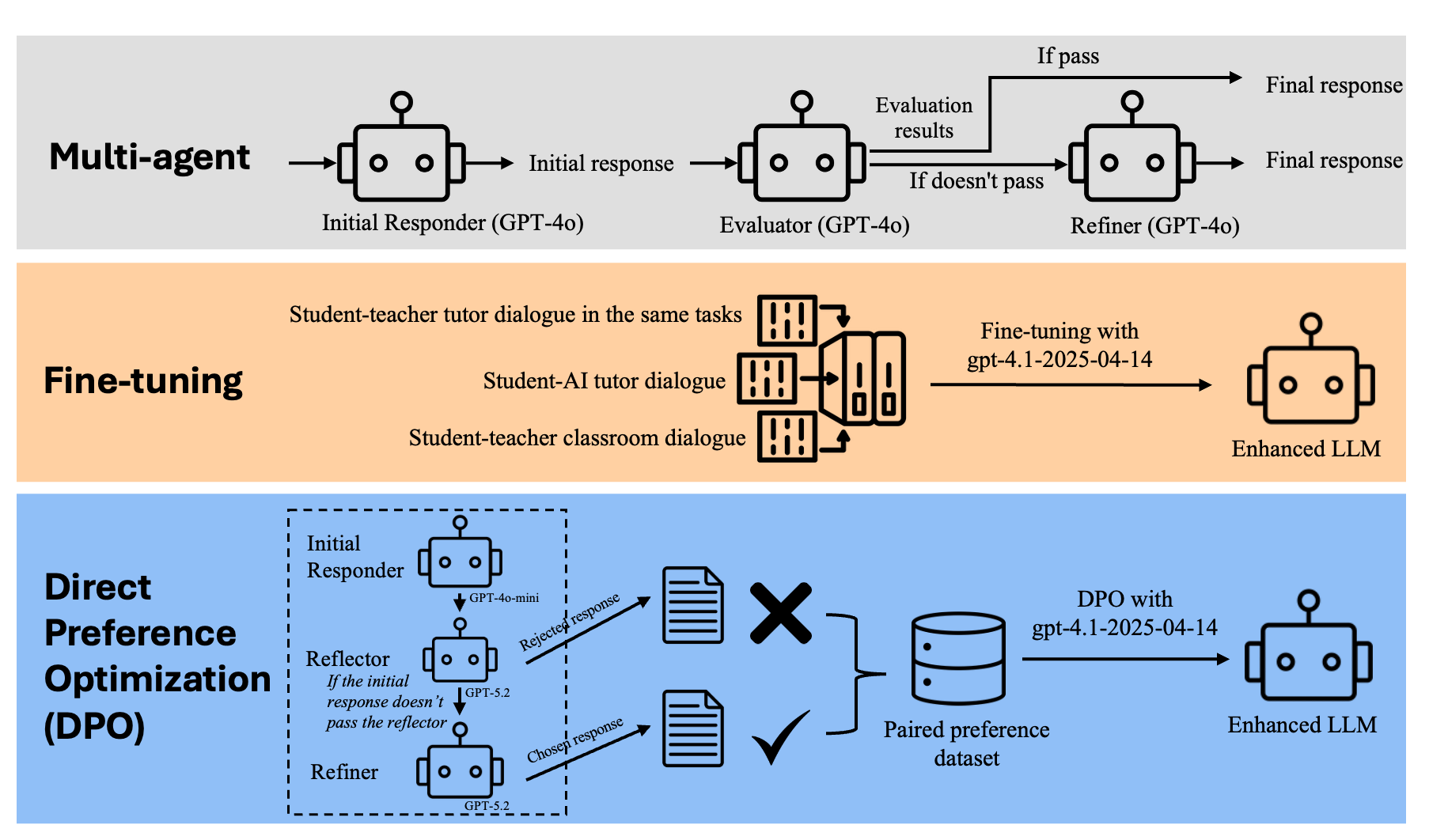}
    \caption{Overview of the approaches. LLM selection considering commercial (e.g., GPT) and open-source LLMs (Llama, Mistral) from testing-feedback in Summer-Fall 2025.}
    \label{fig:framework}
\end{figure}

Building on our evaluation (section 3.2), we explored three approaches to improve authenticity: Multi-agent (responder-evaluator-refiner), Fine-tuning (using real-world classroom data), and DPO (training from preference data; Figure  \ref{fig:framework}).

\subsubsection{Multi-agent}
The Multi-agent approach involved decomposing a main objective into specialized tasks for distinct collaborating agents using \verb|gpt-4o| \cite{cao2025first,li2024survey}. We designed three agents (see Appendix B for prompts). An \textit{Initial Responder} outputted the responses based on the student's profile specifications. Then, an \textit{Evaluator} agent used the evaluation framework for Cognition and Language (section 3.2) to determine the responses' authenticity and explain its judgment. Finally, a \textit{Refiner} agent revised the responses according to the Evaluator agent's feedback. All agents had access to the chat history as part of the input.

\subsubsection{Fine-tuning}
We fine-tuned \verb|gpt-4.1-2025-04-14| via OpenAI’s API (epochs=3, batch size=4, learning rate multiplier=0.2) with a 90/10 train-test split. To capture task-specific and general mathematical discourse, our training data included 1,296 student utterances from (1) previous PST-student interactions from the same fraction task (96 utterances), (2) student-AI tutor chats on fraction tasks from Khan Academy (917 utterances) \cite{miller2024llm}, and (3) 35 transcripts of elementary lessons on fractions (283 student utterances; TalkMoves dataset \cite{suresh2022talkmoves}).

%Fine-tuning represents a supervised learning process wherein a pre-trained model is further trained on labeled datasets, to align its outputs with specific reference standards. We compiled the training data from three sources: (1) 26 PST-student conversations (96 student utterances; 1-2 sentences per utterance) in the same fraction task, which we collected in previous projects to inform the math simulations, (2) 298 conversations (917 student utterances) from a public Khan Academy dataset featuring interactions between students and an AI tutor focused on fraction \cite{miller2024llm}, and (3) 70 teacher-student, one-on-one interactions (283 student utterances) from the TalkMoves dataset \cite{suresh2022talkmoves}. To align the data from the third source with our context, we selected 35 lesson transcripts on fraction in elementary mathematics. Together, these sources yielded a total of 1296 student utterances, enabling fine-tuning that captured both task-specific exchanges and the broader patterns of student mathematical discourse. We randomly partitioned the data (90\% split for training, 10\% for testing). We utilized OpenAI’s fine-tuning API and \texttt{gpt-4.1-2025-04-14} (the latest model for fine-tuning) as the base model,  $\texttt{n\_epochs}=3$, $\texttt{batch\_size}=4$, and $\texttt{learning\_rate\_multiplier}=0.2$.

\subsubsection{DPO}
\textbf{Automated preference data construction.} To overcome the scarcity of natural paired responses, we used a Reflexion loop \cite{shinn2023reflexion} to synthesize 150 preference pairs (labeled ``preferred''-``not preferred''). An \textit{Initial Responder} (\texttt{GPT-4o-mini}) generated a baseline output. A \textit{Reflector} agent (\texttt{GPT-5.2}) assessed this using our authenticity framework, prompting a \textit{Refiner} (\texttt{GPT-5.2}) to revise responses that failed the evaluation criteria.
\textbf{Tuning via DPO.} We then trained \texttt{gpt-4.1-2025-04-14} on the preference pairs using DPO, an algorithm that optimized policies to match preferences without RLHF or separate reward models \cite{rafailov2023direct}. We specified \texttt{beta}=0.1 to prioritize the preferences over previous behaviors.

\subsection{Research Procedures}
\subsubsection{RQ1: Authenticity of the Approaches, Compared to Baseline Prompts}
To evaluate whether the three approaches improved authenticity, we used responses marked as inauthentic in the first implementation (\textit{n}=40). We extracted the conversational history (excluding the inauthentic responses) and PSTs' questions as input. Two researchers coded the generated responses (\textit{n}=120) for Cognition and Language authenticity. We used McNemar's test to examine if authenticity differed between the original and revised versions.

Further, we randomly sampled 40 originally authentic interactions, generating 80 responses (40 inauthentic, 40 authentic) per approach (\textit{n}=240 total). We analyzed this full dataset using Generalized Linear Mixed Models (GLMM). Binary authenticity codes (0/1) were the outcome variables, and approach was the predictor. We specified a random intercept for interaction ID to capture shared variance within the same conversational context.

%\subsubsection{Evaluation by coding}
%To evaluate the efficacy of our proposed approaches, we conducted a targeted re-evaluation focusing on specific interaction points from the initial simulation. We first identified a set of 40 dialogue turns where the original student agent's responses were flagged as ``inauthentic'' by domain experts. For each instance, we extracted the preceding conversational history (excluding the problematic response) and utilized it as the input prompt for the three distinct variants of the simulated student agent (Josh). The newly generated substitute responses were then subjected to re-annotation by experts using the established rubric to determine if the authenticity had improved. 
%Furthermore, to establish a comparative baseline and assess model stability, we randomly sampled an additional 40 interaction points where the original responses were classified as ``authentic.'' Following the identical procedure, we utilized the corresponding conversation contexts to generate new responses via the three agent variants. These outputs were subsequently evaluated by the experts.

\subsubsection{RQ2: Teachers' Feedback on Authenticity and Pedagogical Utility}

We evaluated the three approaches using a within-subjects design with eight participants: five PSTs familiar with the baseline agent and three math education researchers (see Appendix C for demographics). Participants completed 45-minute, screen and audio-recorded Zoom interviews and received \$30 for their participation. Each participant interacted with all three agent versions in a randomized, counterbalanced order. On average, the response time was 2.05s (Fine-tuning), 3.37s (DPO), and 4.82s (Multi-agent). 

During the interactions, participants engaged in a fraction task to elicit student thinking (Figure \ref{fig:josh}). Following each interaction with an agent (7–9 minutes), they completed an 11-item usability and effectiveness survey \cite{pan2025tutorup}. Throughout the interviews, we prompted participants to explain their perceptions of the agents and the instructional decisions they made. After completing all interactions, participants ranked the agents for authenticity and preference and shared insights they gathered about student thinking. 
We conducted three rounds of qualitative coding. Two authors separately read the data and generated open codes (in vivo). They then engaged in two discussions to group the codes into broader themes and perform axial coding to synthesize the codes. %Final themes corresponding to authenticity included natural interaction, realistic cognition, and uncertainty. Pedagogical implications included adaptive questioning and reflection on student thinking.

\section{Results}
\subsection{RQ1: Comparative Performance of the Approaches}
\begin{figure}
    \centering
    \includegraphics[width=.45\linewidth]{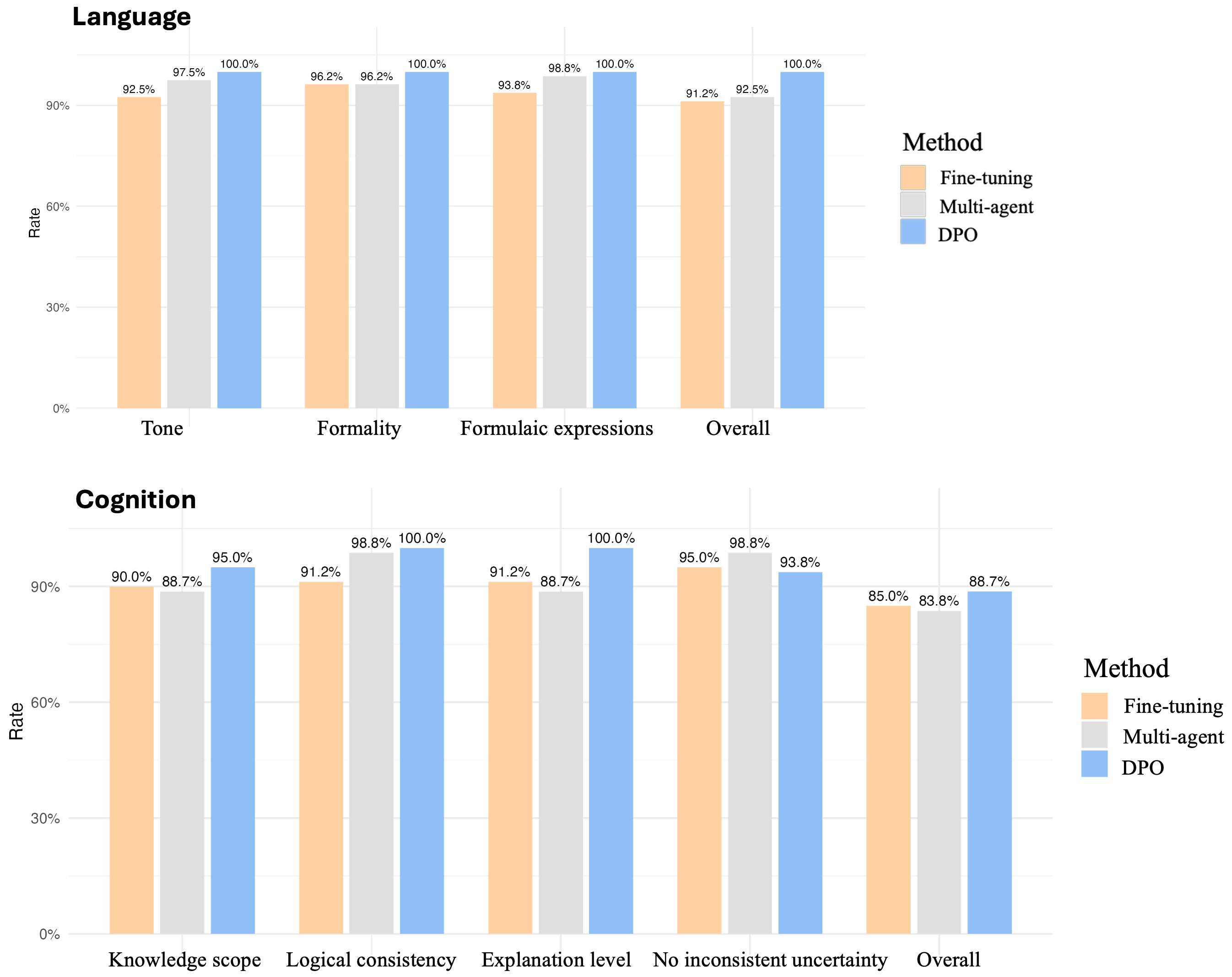}
    \caption{Overall performance of the three approaches: Cognition and Language}
    \label{fig:RQ1}
\end{figure}

McNemar's tests indicated that compared to the baseline few-shot prompts, the three approaches significantly improved Cognition (Fine-tuning: $p=.007$, Multi-agent: $p<.001$, DPO: $p=.013$) and Language (Fine-tuning: $p<.001$, Multi-agent: $p=.003$, DPO: $p<.001$). We also evaluated the combined dataset (baseline inauthentic=50\%). All approaches showed comparatively higher authenticity, with DPO achieving the highest descriptive performance (100\% in language, 88.7\% in cognition; Fig.~\ref{fig:RQ1}). GLMM results revealed no significant difference among approaches for Cognition ($p=.59$) or Language ($p=.91$).
\begin{figure}
    \centering
    \includegraphics[width=0.45\linewidth,trim={0cm 0cm 0cm 2cm}, clip]{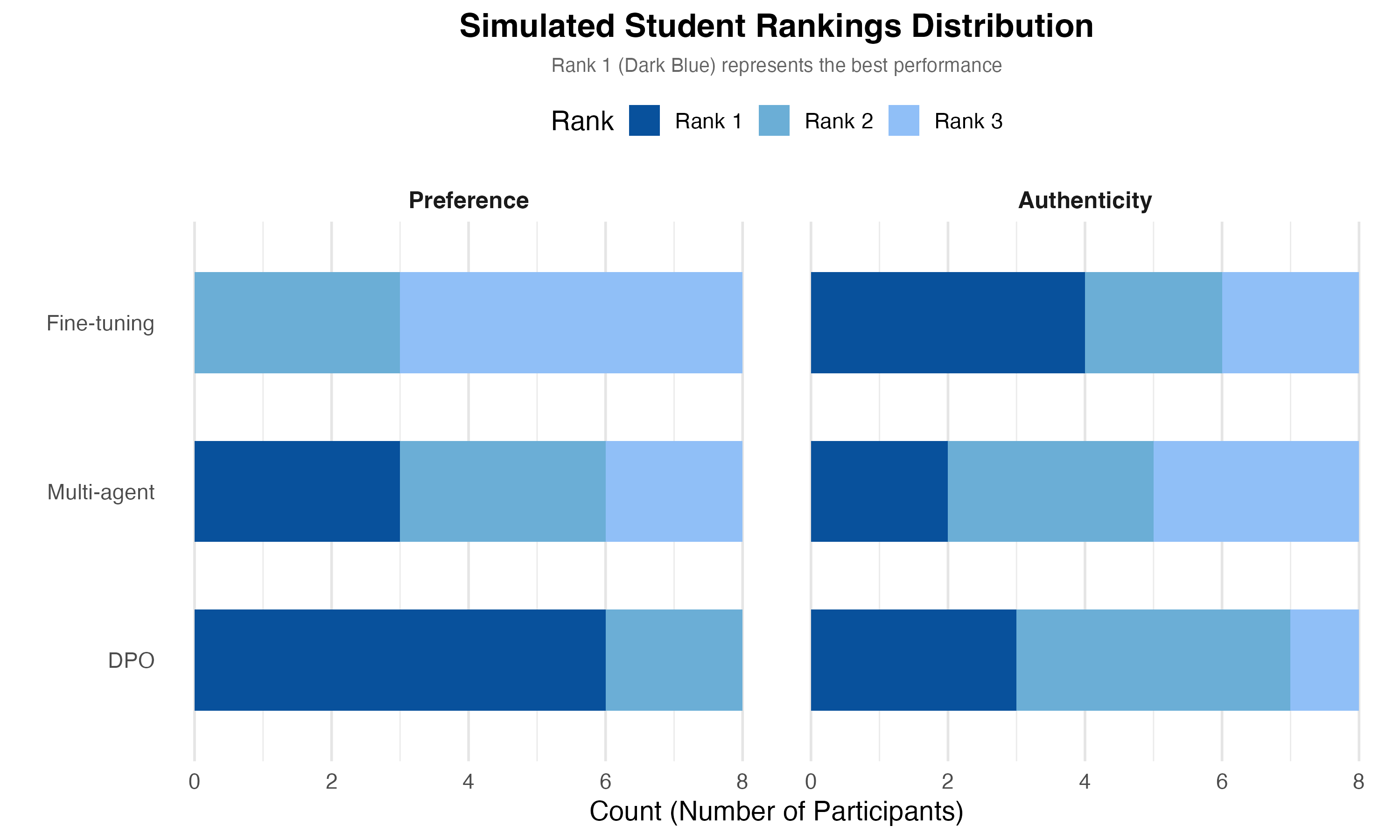}
    \caption{Participants' rankings for preference and authenticity}
    \label{fig:rank}
\end{figure}

\subsection{RQ2: Educators' Feedback: Authenticity \& Pedagogical Utility}
Participants preferred the DPO version (Figure \ref{fig:rank}), rating it higher for relevance, effectiveness, and skill enhancement (see survey results in Appendix D). Perceptions of the agents' authenticity were mixed.

The interview results suggested that participants perceived responses from all approaches as \textbf{authentic but in qualitatively different ways}.  \textit{Natural interactions}. Most participants evaluated that the Fine-tuning approach produced shorter responses that mirrored classroom instances when a student ``isn't super into it'' (Alice, PST) or ``is just waiting for more directions'' (Diana, PST). 

\textit{Cognition}. Participants offered mixed assessments of the Multi-agent version's verbose reasoning. For example, \textbf{Josh (multi-agent)}: ``Hmmm, I tried drawing a number line to find a fraction between 2/3 and 7/8, but I'm not sure where 6/8 fits. [...] I think it's tricky 'cause I'm not sure where to start and end the lines...''
Although three participants noted that this level of reasoning might be expected among higher-performing students, two found it too long and complete. Meanwhile, six participants shared that the DPO approach reflected diverse reasoning. In addition to the prompted strategy to the fraction task (number line), the DPO agent also discussed equivalent fraction. 

\textit{Uncertainty.} Both the Multi-agent and DPO approaches frequently expressed uncertainty (e.g., ``I don't know''). While uncertainty can be productive to probe student understanding, three participants highlighted that uncertainty felt inconsistent, particularly when the agent already knew the answers.

\textbf{Different agent interactions provided pedagogical utility, influencing how teachers asked questions and reflected on student thinking}.  \textit{Adaptive questions}. The response characteristics, including expressed uncertainty, brevity, and reasoning, shaped instructional moves. When the agent expressed strong uncertainty, three participants reported shifting toward direct instruction. Concise responses (Fine-tuning version) prompted open-ended questions to elicit solutions (e.g., ``How did you do it?''). While participants recognized the utility of brief responses, two noted that they could be demotivating, as if ``the student did not want to learn'' (George, PST). Meanwhile, the detailed reasoning of the Multi-agent and DPO versions encouraged deeper ``how'' and ``why'' questions (e.g., ``Why do you think this will help?''). 

\textit{Reflection on student thinking}. All participants shared that the simulations fostered reflection to inform future teaching. Bella (PST) reflected: ``I like that there are 3 different versions because students might respond very differently.'' Edward (PST) noted: ``I refer back to [what I've learned] that talks about when kids shut down, or when they’re in spaces where they’re open to learning. I think about how I present questions to kids to build them up.''

\section{Discussion and Conclusion}
The approaches (Fine-tuning, Multi-agent, and DPO) improve authenticity, compared with the few-shot baseline (RQ1). Each approach has unique affordances. Fine-tuning suits data-rich environments \cite{Barrett2025,zheng2025cognitive}; DPO (via Reflexion \cite{shinn2023reflexion}) overcomes data scarcity; and Multi-agent frameworks require no data but can have high latency. Practically, qualitative findings suggest the need to balance the approach with teachers' practice goals (RQ2). Simulations with more explicit reasoning (e.g., Multi-agent, DPO) can build confidence in interacting with students, while those with uncertainty (Fine-tuning) can prompt teachers to check students' understanding. %Developers can tailor agents to specific training goals: Multi-agent and DPO models (explicit reasoning) build teacher confidence, while Fine-tuned models (shorter, uncertain responses) effectively train teachers to elicit reasoning. 
Future work can integrate knowledge tracing to resolve inconsistent uncertainty; incorporate human preferences and measurement models in DPO \cite{scarlatos2025smart}; involve additional tasks and multi-student interactions; and pair similations with AI or human coaching to deepen professional noticing.

 \bibliographystyle{splncs04}
\bibliography{Ref}





% LNCS Homepage, \url{http://www.springer.com/lncs}, last accessed 2023/10/25
% \end{thebibliography}
\end{document}